\documentclass{rnaastex}

\begin{document}

\title{2MASS J06562998+3002455: Not a Cool White Dwarf Candidate, but a Population II Halo Star} 

\correspondingauthor{Ra\'ul~de~la~Fuente~Marcos}
\email{rauldelafuentemarcos@ucm.es}

\author[0000-0002-5319-5716]{Ra\'ul~de~la~Fuente~Marcos}
\affiliation{AEGORA Research Group,
             Facultad de Ciencias Matem\'aticas,
             Universidad Complutense de Madrid, 
             Ciudad Universitaria, E-28040 Madrid, Spain}

\author[0000-0003-3894-8609]{Carlos~de~la~Fuente~Marcos}
\affiliation{Universidad Complutense de Madrid,
             Ciudad Universitaria, E-28040 Madrid, Spain}


\section{} 

2MASS~J06562998+3002455 or PSS~309-6 is a high proper-motion star that was discovered during a survey with the 2.1~m telescope at Kitt Peak 
National Observatory \citep{1991AJ....102..395H}. The nature of this object was investigated by \citet{2005NewA...11...59D} using rather 
uncertain data to conclude that it could be a nearby cool helium-atmosphere white dwarf candidate; other alternatives discussed by 
\citet{2005NewA...11...59D} included the possibility of PSS~309-6 being a Population II subdwarf ejected from a globular cluster with a
tangential velocity as high as $435_{-85}^{+111}$~km~s$^{-1}$. Here, we reevaluate the status of this interesting star using {\it Gaia} DR2
(source 888147903365642112).

{\it Gaia} DR2 \citep{2016A&A...595A...1G,2018arXiv180409365G} supplies high-quality input data ---namely, equatorial coordinates, parallax, 
radial velocity, proper motions, and their respective standard errors--- to study the kinematics of PSS~309-6, which is placed at a distance 
of $560_{-56}^{+70}$~pc, i.e. it is not located nearby. {\it Gaia} DR2 does not provide a value for the radial velocity of PSS~309-6 and we 
will assume that its radial motion is negligible in comparison with its tangential one. Following the approach outlined by 
\citet{1987AJ.....93..864J}, we have used the input data to compute Galactic space velocities and their uncertainties in a right-handed 
coordinate system for $U$, $V$, and $W$; axes are positive in the directions of the Galactic Center, Galactic rotation, and the North 
Galactic Pole. The necessary values of the Solar motion were taken from \citet{2010MNRAS.403.1829S}. 

\begin{figure}[!ht]
\begin{center}
\includegraphics[scale=0.39,angle=0]{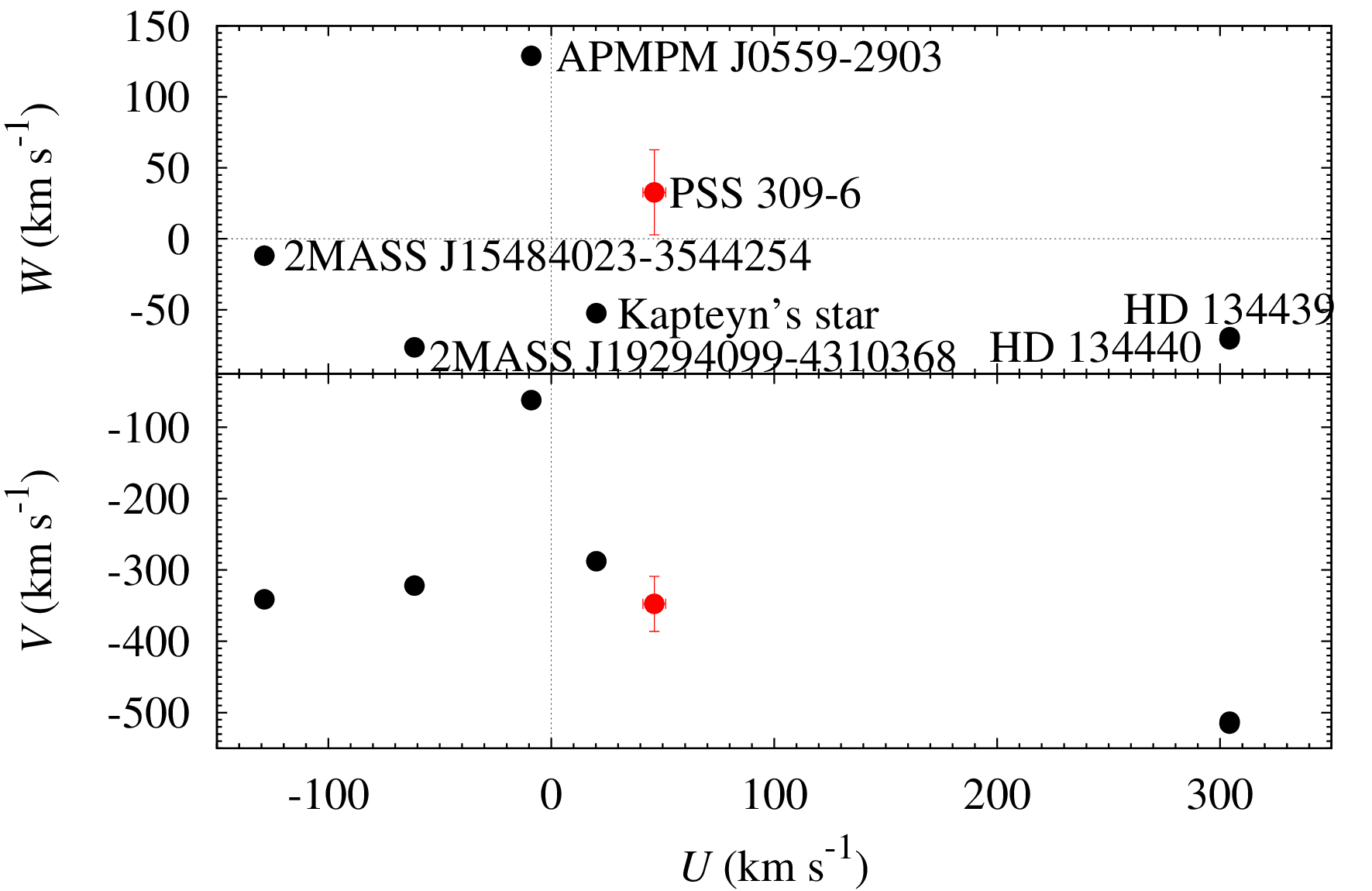}
\includegraphics[scale=0.39,angle=0]{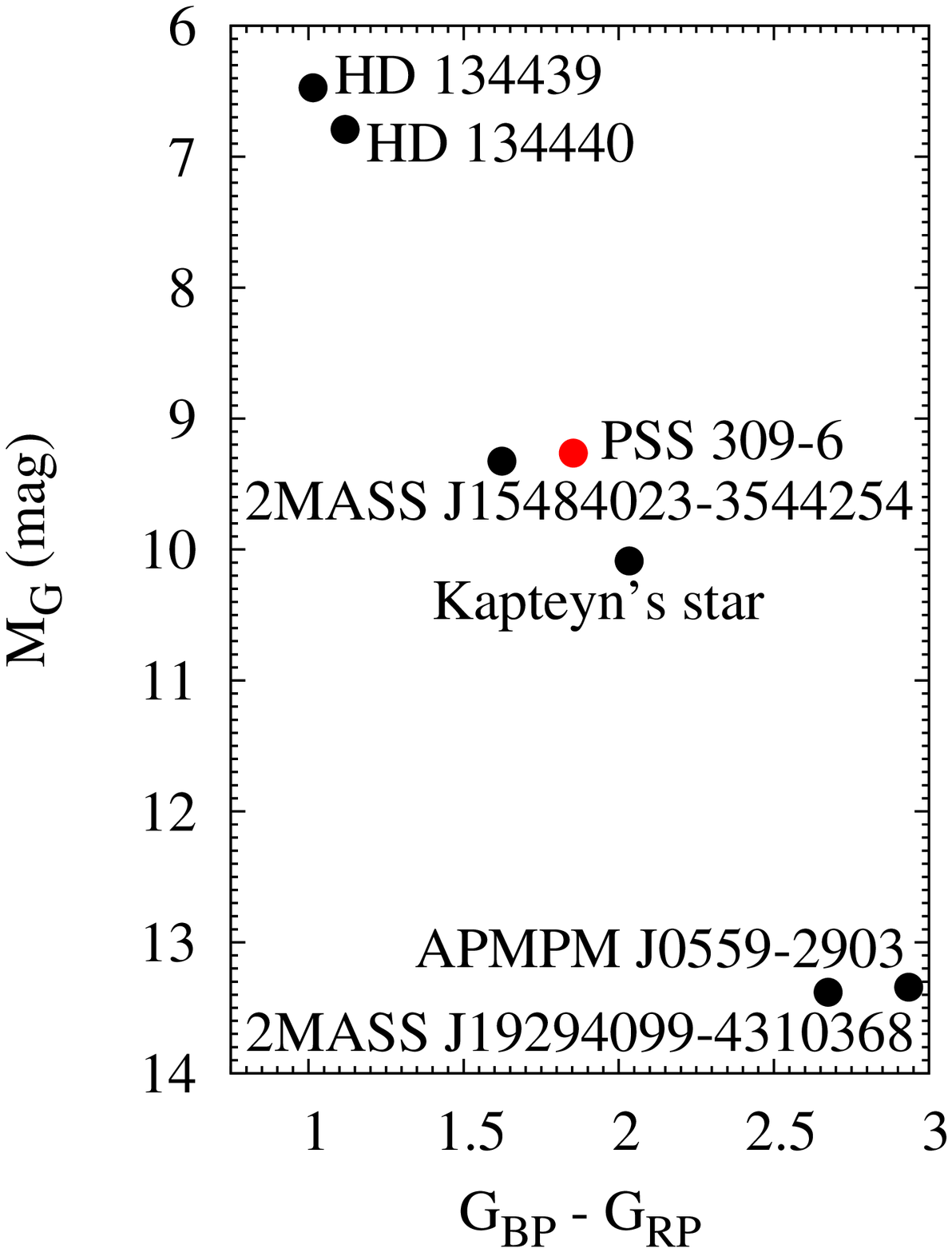}
\caption{The left-hand side panels show the heliocentric Galactic velocity components of PSS~309-6 (in red) and six Population II subdwarfs 
         (in black, HD~134439 and HD~134440 have nearly identical values of $U$, $V$, and $W$). The right-hand side panel displays the 
         Hertzsprung-Russell diagram of the same stars, similar to fig. 5 in \citet{2018arXiv180409378G}. The input data used to prepare 
         this figure are from {\it Gaia} DR2. 
\label{fig:1}}
\end{center}
\end{figure}

Computed as indicated above, the heliocentric Galactic velocity components are $U = 46\pm5$~km~s$^{-1}$, $V = -348\pm39$~km~s$^{-1}$, and 
$W = 33\pm30$~km~s$^{-1}$ (Figure~\ref{fig:1}, left-hand side panels); the corresponding Galactocentric Galactic velocity components are 
$U = -35\pm5$~km~s$^{-1}$, $V = -92\pm39$~km~s$^{-1}$, and $W = 40\pm30$~km~s$^{-1}$. These results strongly suggest that PSS~309-6 could be
a Population II star as the value of its $V$ component is close to $V = -220$~km~s$^{-1}$, which is typical for halo stars in the immediate
solar neighborhood \citep{1998A&A...329...81F}. The average $V$ velocity for the halo is still under debate, but probably lies between 
$-270$ and $-180$~km~s$^{-1}$ \citep{1993ARA&A..31..575M}. PSS~309-6 appears to be a relatively distant Population II subdwarf, not a nearby 
white dwarf as argued by \citet{2005NewA...11...59D}. We have compiled {\it Gaia} DR2 data on six well-studied halo stars of diverse 
spectral types ---Kapteyn's star (sdM1), APMPM~J0559-2903 (sdM7), HD~134439 (sdK1), HD~134440 (sdK2), 2MASS~J15484023-3544254 (sdK5), and 
2MASS J19294099-4310368 (sdM7)--- and plotted them in Figure~\ref{fig:1} to place PSS~309-6 in its proper context. Both the kinematic and 
photometric properties of PSS~309-6 and the subdwarfs are consistent: it could be a late K dwarf/early M subdwarf.
 
Kapteyn's star is the nearest known halo star and PSS~309-6 exhibits similar kinematic and photometric signatures (Figure~\ref{fig:1}).
Its properties also resemble those of 2MASS~J15484023-3544254, which was once thought to be the nearest cool white dwarf but was later
reclassified as K-type subdwarf \citep{2004A&A...425..519S,2005ApJ...627L..41F}. Although it is virtually certain that PSS~309-6 is not a 
nearby white dwarf but a more distant Population II subdwarf, further spectroscopic information, including radial velocity measurements, is 
necessary to fully characterize this probable member of the Galactic halo.

\acknowledgments

We thank J. Farihi for comments on our 2005 results that motivated the present study. This work was partially supported by the Spanish 
MINECO under grant ESP2015-68908-R. In preparation of this Note, we made use of the NASA Astrophysics Data System. This research has made 
use of the SIMBAD database and the VizieR catalogue access tool, operated at CDS, Strasbourg, France. This work has made use of data from 
the European Space Agency (ESA) mission {\it Gaia} (\url{https://www.cosmos.esa.int/gaia}), processed by the {\it Gaia} Data Processing and 
Analysis Consortium (DPAC, \url{https://www.cosmos.esa.int/web/gaia/dpac/consortium}). Funding for the DPAC has been provided by national 
institutions, in particular the institutions participating in the {\it Gaia} Multilateral Agreement.

\end{document}